\documentstyle[preprint,aps]{revtex}
\begin{document}
\draft
\title{Shape Fluctuations of a Droplet\\ Containing a Polymer}
\author{Mark Goulian\footnote{Address after Aug. 31:
{\it Center for Studies in Physics and Biology,
The Rockefeller University, 1230 York Avenue,
New York, NY 10021.}} and Scott
T. Milner}
\address{Exxon Research and Engineering Company\\ Annandale,
New Jersey 08801}
\date{\today}
\maketitle
\begin{abstract}
We consider the problem of an ideal polymer confined in a droplet.
When the droplet radius is smaller than the (unconfined) polymer
radius of gyration, the polymer entropy will depend on the droplet
shape. We compute the resulting surface free energy. Using parameters
appropriate for polymers confined in microemulsions, we find that the
polymer and bending surface energies are comparable for the lowest modes.
Finally, we argue that chain self-avoidance will decrease the strength of
the polymer contribution to the surface energy.
\end{abstract}


The Hamiltonian governing surface shapes usually follows from
considerations of two-dimensional elasticity and
geometry. Thus, fluid surfaces such as
bilayers, emulsions and microemulsions are
well-described by surface tension and curvature elasticity, while solid
membranes also depend on in-plane shear and compression moduli \cite{jur}.
In general, however, one must also take into account the
response of the surrounding three-dimensional medium. (For
convenience we speak of only one medium although in general the surface
may separate two different materials.) The medium is characterized by a
length, $\xi$, which describes the distance over which disturbances are
screened in the bulk. If we consider only surface perturbations with
wavelengths larger than $\xi$, the bulk will contribute to the surface
elastic constants but the form of the Hamiltonian will be as described
above. For perturbations with wavelengths below $\xi$, on the other hand,
the bulk response gives rise to new terms in the surface Hamiltonian.

As an illustration, consider a surface bounding an elastic solid with
modulus $B$. In this case, $\xi$ is infinite. The contribution to the
surface energy from the bulk elasticity can be estimated
by dimensional analysis for surface perturbations with wavelengths much
less than the size of the system. If we allow a surface
tension, $\mu$, and bending rigidity, $\kappa$, then the
energy per area of a surface displacement
$u(q)\exp(iqr)$ must take the form:
\begin{equation}
{\rm Energy/Area}\sim(Bq+\mu q^2+\kappa q^4)|u(q)|^2.
\end{equation}
Thus, for sufficiently small $q$, the bulk elasticity dominates.

In this paper we consider an approximately spherical surface
(droplet) enclosing a single polymer. We are interested in the limit in
which the radius of the droplet is smaller than the radius of
gyration of the unconfined polymer. We were motivated by recent work
in which individual polymers
have been confined within microemulsion droplets \cite{lal},\cite{huang}.
In order to make the problem tractable, we consider ideal polymers, i.e.
we neglect  oidance. We will return to this point at the end.

The droplet is assumed to be incompressible, with volume $V=4\pi
R^3/3$. Points on the droplet surface, $S$, are parametrized
in terms of the radial displacement from the sphere with radius $R$:
\begin{equation}
{\bf R}_S(\theta,\phi)=(R+u(\theta,\phi)){\bf \hat r}.\label{u}
\end{equation}
Incompressibility implies:
\begin{equation}
\int{\rm d}\Omega\;{1\over3}R_S(\theta,\phi)^3={4\over3}\pi R^3.
\label{cnstrnt}
\end{equation}
After expanding $u$ in normalized spherical harmonics \cite{spfnc},
\begin{equation}
u(\theta,\phi)=\sum_{l=0}^\infty\sum_{m=-l}^l
u_{lm}Y_{lm}(\theta,\phi), \label{exp}
\end{equation}
we enforce the constraint Eq.\ (\ref{cnstrnt}) by solving for
$u_{00}$:
\begin{equation}
u_{00}=-{1\over2R\sqrt{\pi}}\sum_{lm}|u_{lm}|^2+O(u_{lm}^3).\label{cnsolve}
\end{equation}

The surface free energy of the droplet has the form
${\cal F}={\cal F}_b+{\cal F}_p$, where ${\cal F}_p$ is the free energy
of the confined polymer and ${\cal F}_b$ is the  bending energy of the
droplet  surface (c.f. \cite{milsaf}):
\begin{eqnarray}
{\cal F}_b={\kappa\over2}\sum_{l>0,m}\,{|u_{lm}|^2\over
R^2}(l+2)(l-1)(l(l+1)-4w+2w^2).
\end{eqnarray}
$\kappa$ is
the bending rigidity and $w$ is the ratio of $R$ to the spontanteous
curvature radius. For simplicity, we have ignored surface tension,
although it may easily be included. ${\cal F}_p$ is the free energy of the
confined polymer.

For an ideal polymer of $N$ segments, ${\cal F}_p$ may be expressed in
terms of the statistical weight $G_N(r_0,r_1)$,
which is the number of walks that start at $r_0$ and end at $r_1$
after N steps:
\begin{equation}
{\cal F}_p=-k_BT\log\left[\int_V{\rm d}^3r_0\int_V{\rm
d}^3r_1\;G_N(r_0,r_1)\right].\label{pfunc}
\end{equation}
$G_N({\bf r_0},{\bf r_1})$ satisfies the diffusion equation (c.f.
\cite{dg})
\begin{equation}
{\partial\over\partial
N}G_N({\bf r_0},{\bf r_1})={b^2\over6}\nabla_{r_1}^2G_N({\bf r_0},{\bf
r_1}),
\end{equation}
where $b$ is the effective bond length.
The radius of gyration of the polymer, $R_g$, is related to $N$ by
$R_g^2=Nb^2/6$.
For either ${\bf r_0}$ or ${\bf r_1}$ on the boundary of the droplet, $S$,
$G_N({\bf r_0},{\bf r_1})$ vanishes.
In terms of orthonormal eigenfunctions $\psi_i$ of the Laplacian,
satisfying
\begin{equation}
\nabla^2\psi_i=-k_i^2\psi_i\hskip.5cm\psi_i({\bf r})|_{{\bf r}\in S}=0,
\label{eigen}
\end{equation}
we have the expansion
\begin{equation}
G_N({\bf r_0},{\bf r_1})=\sum_i \psi_i^*({\bf r_0})\psi_i({\bf
r_1})\exp(-k_i^2 R_g^2).\label{plsum}
\end{equation}
Since $S$ is a perturbation of a sphere of radius $R$, Eq. (\ref{u}), the
eigenvalues $k_i^2$ are proportional to $1/R^2$. In the limit $R\gg R_g$,
$G_N({\bf r_0},{\bf r_1})$ falls off rapidly for $|{\bf r_0}-{\bf
r_1}|\gg R_g$, and Eq. (\ref{pfunc}) reduces to the translational free
energy of the polymer, as expected: ${\cal F}_p\approx -k_BT\,\log(V)$.

For $R_g>R$, the sum in Eq.\ (\ref{plsum}) can be approximated by
keeping only the lowest eigenvalue $k_0^2$ (ground-state dominance
\cite{dg}). Within this approximation, the free energy is given by
\begin{eqnarray}
{\cal F}_p&=&k_BT R_g^2k_0^2 -
2k_BT\log(c_0)+
O\left(e^{-R_g^2(k_1^2-k_0^2)}\right)\nonumber\\
c_0&\equiv&\int_V{\rm d}^3r\;\psi_0(r).\label{free1}
\end{eqnarray}
The first term on the right-hand side of Eq. (\ref{free1}) is the leading
term (in $R/R_g$). We have also included the first correction (the second
term in Eq. (\ref{free1})) because it may be calculated with little extra
work (see appendix). Note that the remaining higher order corrections are
exponentially suppressed. Thus, for a confined chain, the shape dependence
of ${\cal F}_p$ is determined by the shift in $k_0^2$ and $\psi_0$ for
perturbations of the form Eq. (\ref{u}).

As a simple application of the above formula, we first compute the
free energy of the polymer as a function of the droplet radius.
We therefore take $R_S=R$ and the eigenvalues in Eq. (\ref{eigen}) reduce
to $(n\pi/R)^2,\ n=1,\ldots$. Thus, for $R<R_g$, the
free energy is given by
\begin{equation}
{\cal F}_p\approx k_BT\pi^2{R_g^2\over R^2}.\label{fsph}
\end{equation}

It is tempting to view the interior of the droplet as an elastic solid with
bulk modulus $B$. To estimate $B$, we relax the requirement of
incompressibility and compute the change in free energy under a uniform
change in the radius of the droplet. We thus take $R\to R-u$ and expand
Eq. (\ref{fsph}) in the strain $u/R$; the term linear in $u/R$ simply
reflects the fact that we are perturbing about a compressed chain.
Dividing by the droplet volume gives a bulk modulus of order $B\sim
k_BT\,R_g^2/R^5$. We can now use the result from the introduction to
estimate ${\cal F}_p$ for surface disturbances with wavelength much
smaller than $R$. This corresponds to a mode $u_{lm}$ with
wavevector $q\sim l/R$, $l\gg1$. From the argument in the introduction
we then find
\begin{equation}
{\cal F}_p\approx k_BT\,l\,{R_g^2\over R^2}{u_{lm}^2\over
R^2}.\label{largel}
\end{equation}

As we shall see below, ${\cal F}_p$ does indeed take this form in the
limit of large $l$. However, it is in
the low $q$ modes where the contribution from the polymer is expected to
be important (compared with the contribution from the bending energy).
This corresponds to small $l$ (i.e. wavelengths comparable to $R$)
and requires a more careful analysis, which we now present.

As described above, we must find the ground state of Eq. (\ref{eigen}).
The general problem of the
change in  the spectrum of the Laplacian under a boundary perturbation was
considered some time ago and appears in a number of contexts, such as
acoustics in or around irregular regions, cavity resonators, and the
Wigner-Seitz approximation for electrons in a crystal (see \cite{mf} and
references therein). The early work on this problem focused on the
solution to first order in the perturbation. Unfortunately, we require
the result to second order (i.e. $O(u_{lm}^2)$), which is substantially
more difficult to determine. In \cite{mf} a general formula
is given for the change in the eigenvalues to second order, however
the final formula appears to be incorrect. Therefore, in the
appendix we calculate the ground-state eigenvalue and integral of the
ground-state wavefunction for a perturbation of the form Eq. (\ref{u}).

Combining the results of the appendix
with the constraint Eq. (\ref{cnsolve}) gives:
\begin{eqnarray}
k_0^2&=&{\pi^2\over R^2}\Big[1+\sum_{l>0,m}\Big({l+2\over2\pi}-
{f_l\over2}\Big){|u_{lm}|^2\over R^2}+
O(u_{lm}^3)\Big]\label{gst}\\
c_0&=&2\sqrt{2\over\pi}R^{3/2}\Big[1+{1\over8}\sum_{l>0,m}
\Big(2(l+2)f_l-\pi f_l^2 \label{igst}
-3{(l+2)\over\pi}\Big){|u_{lm}|^2\over R^2}+O(u_{lm}^3)\Big],\nonumber
\end{eqnarray}
where $f_l\equiv{j_{l+1}(\pi)/j_l(\pi)}$ and $j_l$ is the spherical
Bessel function of order $l$. Note that the \hbox{$u_{1m}$-dependent}
terms in the volume preserving perturbation Eqs. (\ref{u},\ref{cnsolve})
correspond to infinitesimal translations, which should not change  the
spectrum of the Laplacian. One may readily check that indeed the order
$u_{1m}^2$ terms in Eqs. (\ref{gst}-\ref{igst}) vanish.

Substituting the above results into Eq. (\ref{free1}), and dropping an
additive constant that is independent of $u_{lm}$, we find for the
polymer contribution to the droplet free energy:
\begin{eqnarray}
{\cal F}_p&=&{k_BT\over2}\sum_{l>0,m}{|u_{lm}|^2\over R^2}\,\,\Gamma_l +
\ldots\\ \Gamma_l&\equiv&\Big[{R_g^2\over R^2}\pi(l+2-\pi f_l)
+{\pi\over4}f_l^2-{l+2\over2}f_l+{3l+6\over4\pi}\Big].\nonumber
\end{eqnarray}

In the limit of large $l,$ $\Gamma_l$ has the behavior
\begin{equation}
\Gamma_l\mathop{\sim}\limits_{l\to\infty}l\,\pi({R_g^2\over
R^2}+{3\over4}).
\end{equation}
We thus recover the form predicted from simple elasticity
Eq. (\ref{largel}).

As discussed above, the $l=0$ mode is constrained by
incompressibility and the $l=1$ modes correspond to translations
(which implies $\Gamma_1=0$). Therefore, the leading contribution in the
small-$l$ or long-wavelength limit comes from $l=2$:
\begin{equation}
\Gamma_2={R_g^2\over R^2}{\pi\over3}\big(\pi^2-3)+
{(\pi^2-9)(\pi^2+3)\over36\pi}\approx7.2{R_g^2\over R^2}+0.1.
\end{equation}

We thus find that for $R_g>R$, the surface free energy picks up a
contribution from the polymer which does not appear to have a
simple interpretation in terms of surface elasticity and geometry.
Furthermore, for the lowest modes, this contribution to the surface energy
can be quite important. For the experiments in \cite{huang}, $R_g/R\approx
4$, $\kappa\approx2k_BT$, and $w\approx2$. With these values, the ratio of
the ($l=2$) polymer contribution to the free energy to the bending
contribution is ${\cal F}_p|_{l=2}/{\cal F}_b|_{l=2}\approx
2.4$; the contributions from bending energy and the polymer are
comparable.

In the calculations above we have neglected self-avoidance. However, in the
limit of very high compressions self-avoidance will certainly be
important. $R$ is bounded from below by $R_m$, the radius
for which the polymer volume fraction is equal to one. Near this
limit, we  expect the interior of the droplet to be similar to a melt.
The screening length $\xi$ \cite{dg} is then microscopic and
the polymer free energy is insensitive to the shape of the droplet:
$\Gamma_l\approx0$. Therefore, for the polymer to make a significant
contribution to the surface free energy, the droplet radius must be in the
range $R_m<R<R_g$. However, even for droplets within this range, self
avoidance will be important for the lowest modes. A simple
scaling estimate  of $\xi$ \cite{dg} gives
\begin{equation}
\xi\approx R\left({R\over R_g}\right)^{1/(3\nu-1)},
\end{equation}
where $R_g\sim N^\nu$. For a chain in a good solvent, $\nu\approx3/5$, and
in a theta solvent, $\nu=1/2$ \cite{dg}.
Thus, even for a modest compression of the polymer
($R \alt R_g$), the screening length is comparable to $R$ ($\xi \alt R$)
and the lowest modes of the droplet are just at the wavelengths
where screening becomes important. While this results in a suppression
of the free energy of the lowest modes, we cannot gauge the size of the
effect. In particular, we do not know whether this suppression is
strong enough to significantly reduce the polymer contribution
to the surface energy relative to the bending energy.

Even for the higher modes, for which
screening is unimportant (i.e. $l>R/\xi$), there
will be corrections due to self-avoidance. We can again use a
scaling argument to estimate the free energy of a polymer in a spherical
droplet (with $R<R_g$):
\begin{equation}
{\cal F}_p\approx k_BT\left(R_g\over R\right)^{1/\nu}.
\end{equation}
Ignoring numerical factors, the above estimate is smaller than the ideal
polymer result, Eq. (\ref{fsph}), by a factor of $(R_g/R)^{2-1/\nu}$.
Following the arguments leading to Eq. (\ref{largel}), we then expect
the large-$l$ behavior of $\Gamma_l$ to be smaller by a factor of
$(R_g/R)^{1/3}$ for a polymer in a good solvent; there is no correction
for a polymer in a theta solvent.

To summarize,
we have determined the contribution to the surface free energy arising
from confining an ideal polymer within a droplet. Not surprisingly,
this contribution does not have a simple interpretation in terms of
two-dimensional elasticity  and geometry. Using parameters appropriate
for polymers confined in microemulsions \cite{huang}, we find
that the bending and polymer terms in the surface energy are
comparable for the lowest modes. Scaling arguments suggest that
chain self-avoidance will tend to decrease the magnitude of the
polymer  contribution.

\noindent{\it Acknowledgments:} We would like to thank J. Huang for useful
discussions.

\appendix
\section*{Boundary Perturbations}
In this appendix we find the ground-state eigenvalue and integral of the
ground-state eigenfunction to second order in the perturbation of the
shape of the boundary, Eq. (\ref{u}). In \cite{mf} a formula for
the eigenvalues of the Laplacian to second order in a boundary
perturbation is given for the case of Dirichlet boundary conditions (Eq.
(9.2.71) of \cite{mf}). We have not been able to reproduce
the derivation of this formula. Furthermore, when one applies the
formula of \cite{mf} to the simple case of a spherical boundary with a
perturbation of the radius, $R\to R-\delta R$, it does not give the
correct result. However, by a similar procedure to that described in
\cite{mf}, we can derive the correct formula for general
perturbations of a sphere.

We follow the notation in \cite{mf} and
denote the perturbed volume by $V$ and the boundary by $S$. $S$ is a
perturbation of $S_0$, which is a sphere of radius $R$. For the purposes
of the derivation, however, we must formally consider perturbations that
are entirely contained within the unperturbed surface. We therefore
consider a second sphere $S'$ with a slightly larger radius than $S_0$,
such that $S$ is entirely contained in $S'$  (Fig. \ref{pertfig}); the
volume contained in $S'$ is $V'$. We will compute the spectrum for the
Laplacian on $V$ in terms of the spectrum on $V'$.  The surface $S$ is
thus given by: \begin{equation}
{\bf r}_S=(R'+v(\theta,\phi)){\bf \hat
r}\hskip1cm v(\theta,\phi)\le0,
\end{equation}
where in terms of the perturbation $u(\theta,\phi)$ ( Eq. (\ref{u}))
\begin{eqnarray}
v(\theta,\phi)&=&u(\theta,\phi)-a\hskip.5cmR'=R+a\nonumber\\
a&\equiv&\mathop{\rm max}\limits_{\theta,\phi}u(\theta,\phi).\label{v}
\end{eqnarray}

We wish to determine the ground-state eigenfunction and eigenvalue on $V$:
\begin{equation}
(\nabla^2+k^2)\psi=0\hskip.25in\psi(r)|_{r\in S}=0.
\end{equation}
in terms of the
orthonormal eigenfunctions and eigenvalues on $V'$:
\begin{equation}
(\nabla^2+k_{nl}^2)\phi_{nlm}=0\hskip.25in
\phi_{nlm}(r)|_{r\in S'}=0.\label{unpert}
\end{equation}
Since $S'$ is a sphere of radius $R'$, Eq. (\ref{unpert}) is easily
solved:
\begin{eqnarray}
&\phi&_{nlm}(r)=\sqrt{2\over R'^3}{j_l(k_{nl}r)\over
j_{l+1}(k_{nl}R')}Y_{lm}(\theta,\phi)\nonumber\\
&k&_{nl}={x_{nl}\over R'}
\end{eqnarray}
where $l=0,1,\ldots,\ m=-l,\ldots l,\ n=1,2,\ldots,$
$j_l(x)$ is the spherical Bessel function of order $l$ and $x_{nl}$ is
the $n$th root of $j_l(x)$ \cite{spfnc}. To leading order we have
$\psi(r)=\phi_{100}(r)+O(v)$ and $k^2=k_{10}^2+O(v)$.
A simple application of Green's theorem
gives:
\begin{equation}
(k^2-k_{10}^2)\int_V{\rm d}^3r\;\phi_{100}(r)\psi(r)=-\oint_S{\rm d}^2r
\;\phi_{100}(r)\left({\partial\psi(r)
\over\partial n}\right),\label{green}
\end{equation}
where we have used the fact that $\psi(r)$ vanishes on $S$;
$\partial/\partial n$ denotes the outward normal derivative at the
boundary. As in \cite{mf}, we solve this integral equation
iteratively in powers of $v$; the right-hand side is $O(v)$, since
$\phi_{100}(r)$ vanishes on $S'$.

At the lowest order, we substitute $\phi_{100}(r)$ for
$\psi(r)$ in Eq. (\ref{green}), express the integrals over $V$ and $S$ in
terms of the nearby $V'$ and $S'$, and expand in $v$ to give
\begin{equation}
k^2-k_{10}^2={2\over2R'\sqrt{\pi}}k_{10}\sum_{lm}k_{1l}\,v_{lm}+O(v^2),
\end{equation} where $v_{lm}$ are defined by expanding $v(\theta,\phi)$
in spherical harmonics:
\begin{equation}
v(\theta,\phi)=\sum_{lm}v_{lm}Y_{lm}(\theta,\phi).
\end{equation}

By a further application of Green's theorem, and
again using Eq. (\ref{green}),
the ground-state eigenfunction may be written as (see \cite{mf})
\begin{equation}
\psi(r)=\phi_{100}(r)+{k_{10}\over2R'\sqrt{\pi}}{\sum_{nlm}}^\prime v_{lm}
{k_{nl}\over k_{nl}^2-k_{10}^2}\phi_{nlm}(r)+O(v^2).\label{psi1}
\end{equation}
The prime on the sum in Eq. (\ref{psi1}) indicates that the
state ($n=1,\ l=m=0$) is excluded. In order to compute $k^2$
to second order in $v$, we would like to substitute the above expression
for $\psi(r)$ into Eq. (\ref{green}). Unfortunately, the sum over $n$ in
Eq. (\ref{psi1}) converges quite slowly ($k_{nl}\sim n$ for large $n$)
and we cannot move the derivative in Eq. (\ref{green}) past this sum.
In order to improve the convergence, we sum the leading large-$n$ behavior
in Eq. (\ref{psi1}) using\footnote{We note that the
identity Eq. (\ref{zero}) is a counterexample to a claimed general result
used in \cite{mf} to improve the convergence of sums arising in boundary
perturbations (see p. 1044 of \cite{mf}).}
\begin{equation}
\sum_{nlm} v_{lm}{\phi_{nlm}(r)
\over k_{nl}} =\sum_{lm}{v_{lm}Y_{lm}\over\sqrt{2R'}}\left({r\over
R'}\right)^l,\label{zero}
\end{equation}
which follows from the identity (for $r<1$) \cite{Wats}:
\begin{equation}
\sum_{n=1}^\infty{j_l(x_{nl}r)\over x_{nl}j_{l+1}(x_{nl})}={r^l\over2}.
\end{equation}

After summing the large-$n$ behavior, Eq. (\ref{psi1}) gives:
\begin{eqnarray}
\psi(r)=\phi_{100}(r)+{k_{10}\over\sqrt{\pi}R'}\Big[
{\sum_{nlm}}^\prime &v_{lm}&{k_{10}^2\over
k_{nl}(k_{nl}^2-k_{10}^2)}\phi_{nlm}(r)\\
&+&\sum_{lm}{v_{lm}Y_{lm}(\theta,\phi)\over\sqrt{2R'}}\left({r\over
R'}\right)^l -v_{00}{\phi_{100}(r)\over k_{10}}\Big] +O(v^2).\nonumber
\end{eqnarray}

This expression converges sufficiently rapidly that it
may be substituted into Eq. (\ref{green}). We
evaluate the resulting sums over $x_{nl}$ by contour integration
\cite{Wats} and then use Eq. (\ref{v}) to express the perturbation in
terms of $u$.  The final result for  the ground state eigenvalue is:
\begin{equation}
k^2={\pi^2\over R^2}\Big[1-{u_{00}\over\sqrt{\pi}R}
+{3\over4\pi}{u_{00}^2\over R^2}+2\pi\sum_{l>0,m}{|u_{lm}|^2\over R^2}
(l+1-\pi f_l)\Big]+O(u^3),
\end{equation}
where $f_l\equiv{j_{l+1}(\pi)/j_l(\pi)}$.

By a similar iteration, again using Eq. (\ref{zero}) to improve
convergence, we find for the integral of the normalized ground-state
eigenfunction:
\begin{eqnarray}
{\int_{V}{\rm d}^3r\,\psi(r)\over\int_{V}{\rm d}^3r\,\psi(r)^2}=
2\sqrt{2\over\pi}R^{3/2}\Big[&1&+{3\over4\sqrt{\pi}}{u_{00}\over
R}+{3\over32\pi}{u_{00}^2\over R^2}\\
&+&{1\over8}\sum_{l>0,m}\Big(2(l+2)f_{l} -\pi
f_{l}^2-3{(l+1)\over\pi}\Big){|u_{lm}|^2\over
R^2}+O(u_{lm}^3)\Big]. \nonumber
\end{eqnarray}


\begin{figure}
\caption{The surface of the droplet, $S$, is a perturbation of a sphere
$S_0$. $S$ may be viewed as a perturbation of a slightly larger sphere,
$S'$, which contains $S$.} \label{pertfig}
\end{figure}

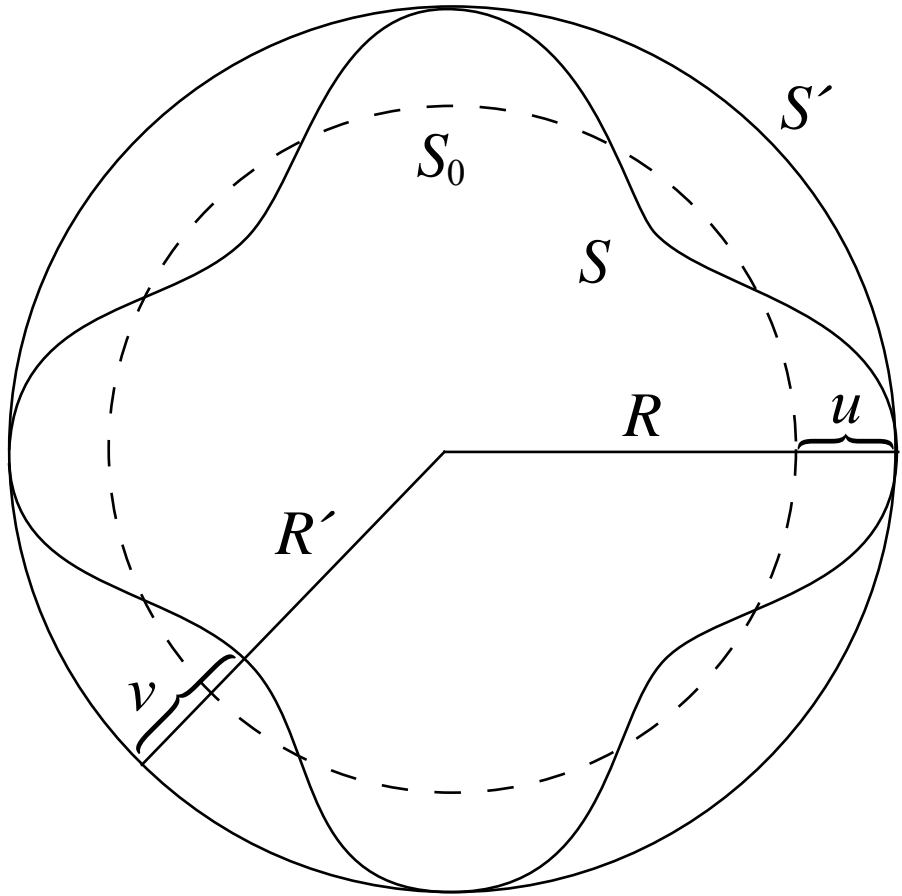
\end{document}